\documentclass[12pt,preprint]{aastex}

\usepackage{amsmath}

\usepackage{epsfig,rotate,graphicx}
\newcommand{\be}{\begin{equation}}
\newcommand{\ee}{\end{equation}}

\begin{document}

\shorttitle{Three body resonances in close orbiting  planetary systems}
\shortauthors{J.C.B. Papaloizou}

\title{Three body resonances in close orbiting  planetary systems:
Tidal dissipation and orbital evolution}

\author{John C. B. Papaloizou}
\affil{Department of Applied Mathematics and Theoretical Physics,\\
University of Cambridge}
\affil{ DAMTP, Centre for Mathematical Sciences,\\
 Wilberforce  Road,\\
 Cambridge,\\
 CB3 0WA 
}
\email {J.C.B.Papaloizou@damtp.cam.ac.uk}  


\begin{abstract}
{We study the orbital evolution of a  three planet system with masses in the super-Earth regime   resulting from  the action of  tides on the planets 
 induced by the central star  which cause  orbital circularization.  We consider systems either in or near to a three body
 commensurability for which adjacent pairs of planets are in a  first order commensurability.
 We  develop a simple analytic solution,   derived from  a time averaged set of equations, that describes the expansion
 of the system  away from  strict commensurability as  a function of  time, once a state where relevant resonant angles undergo small amplitude
librations has been attained.    
We perform numerical simulations that show  the attainment of  such resonant states focusing on the Kepler 60 system.  The 
results of the simulations confirm many of the  scalings
predicted by the appropriate analytic solution.  
 We go on to indicate how the results can be applied to put constraints on the amount of tidal dissipation
that has occurred in the system. 
 For  example,  if  the system has been in a librating state   since its formation,
 we find that its present period ratios imply an upper limit on the time average  of $1/Q',$  with
 $Q'$ being   the tidal dissipation parameter.
On the other hand  if a librating state has not been attained, a lower upper bound  applies.}


\end{abstract}



\section{Introduction}\label{sec1}

The Kepler mission has discovered an abundance of  confirmed and candidate  planets 
orbiting close to their host stars ( Batalha et al. 2013).
Many of these are in tightly packed  planetary systems.
A significant number of these contain pairs that are  close to first order resonances. 
 Lissauer et al. (2011) found  Kepler  candidates in short period orbits in  multiresonant configurations.
The four planet system  KOI-730   exhibits  the mean motion ratios  8:6:4:3  and   
 KOI-500 is a (~near-~) resonant five-candidate system with two three body  mean motion resonances 
 $2n_2- 5n_3+3n_4\sim 0$ and     $2n_3-6n_4+4n_5\sim 0,$ with $n_i$ being the mean motion of planet $i.$  
 More recently Steffen et al. (2013) confirmed the Kepler 60 system  which has 
three planets with the inner pair exhibiting a 5:4 commensurability and the outer pair a 4:3 commensurability.
The case for which the period ratio between consecutive pairs  is 2:1 is termed a Laplace resonance.
 This configuration occurs in the solar system for the  Galilean satellites Io, Europa and Ganymede.
 The configuration of the Kepler 60 system can be regarded as a generalization of this  case  to the situation
 where  the 2:1 resonances are replaced by alternative   first order resonances.

Tightly packed resonant planetary systems are of interest  on account of what their dynamics can tell us
about their origin. In particular the formation of resonant chains is believed to require convergent
migration induced by interaction with the protoplanetary disk  (eg. Cresswell \& Nelson 2006).
In addition tidal interaction with the central star leading to orbital circularization
can be significant in producing tidal heating and  
determining post formation evolution (Papaloizou 2011; Batygin \& Morbidelli 2012; Lithwick \& Wu 2012).
From this it is potentially possible to estimate tidal quality factors which are in turn related to the 
composition and state of the planetary interiors.
We remark that systems with observed  adjacent pairs of first order resonances  may have avoided a form of  long term orbital instability,
associated with higher order many body resonances, that could  cause evolution away from them  in other systems ( Migaszewski et al. 2012).

Migration due to tidal interaction with the disk is a possible mechanism through which
planets end up on short period orbits, as in situ formation implies very massive disks (e.g., Ward 1997;
Raymond et al. 2008). 
 When several protoplanets   in the super-Earh mass range are considered, resonant chains are readily produced
that contain multi-body  resonances (Cresswell \& Nelson 2006). 
Terquem and Papaloizou (2007) 
adopted  a scenario for forming hot super-Earths in which a population of cores that formed
at some distance from the central star migrated inwards due to interaction with the disk.
These collided and merged as they went. This process could produce planetary systems located inside an assumed disk inner edge,  on short period
orbits with mean motions of neighbouring planets that frequently exhibited near commensurabilities.
Note that the extent of the radial migration is not specified  and does not need to be
large in order to produce commensurabilities.
Later evolution due to  circularization of the orbits  induced by tidal  interaction with the
central star, together with later close scatterings and mergers,    caused  the system
to move away from  commensurabilities to an extent determined by the
effectiveness of these processes.

 The  GJ 876  system, which contains a 1:2:4    Laplace resonance, has been studied by   (eg.  Correia et al.  2010;  Gerlach \& Haghighipour 2012; Marti et al. 2013).
Gerlach \& Haghighipour ( 2012) give a detailed  account of the history of the system and show that it is dynamically full.
 Marti et al. (2013) remark that such  a resonant system occupies  small  islands of stability.
Papaloizou and Terquem (2010) considered the system around HD 40307  (Mayor et al. 2009) for which the
pairs consisting of the innermost and middle planets and the middle and outermost planets
are near but not very close to a pair of 2:1 resonances. In spite of this it was found that secular
effects produced by the action of the resonant angles coupled with the action of tides from the
central star could cause the system to increasingly separate from commensurability. 

In this paper we study the evolution of a  three planet system with masses in the super-Earth regime   that is close to a three body resonance under the influence
of tidal effects. We  develop a simple analytic model that describes the evolution of the system,  derived from  a time averaged set of equations,
that describes the increasing  departure of the system  from commensurability with time once a state where resonant angles undergo small amplitude
librations has been attained.    
We give  analytic expressions from which this evolution can be obtained for pairs of general first order resonances when either four resonant angles
or three resonant angles are in a state of small amplitude libration. 
This can be regarded as a  generalisation of  the treatment of two planet resonances given in Papaloizou (2011).

We perform numerical simulations to study  how  a three body  resonance is attained 
 as a result of the operation of circularization  tides  as well as confirm many of the  scalings.
Unlike in Papaloizou \& Terquem (2010), we do not suppose that  the initial system is  already in a
three body resonance as a result of disk induced migration,  but start with the initial orbital configuration reported for the Kepler 60 system
as an example to provide focus. We go on to indicate how the results can be applied to put constraints on the amount of tidal dissipation
that has occurred in the system.


The plan of the paper is as follows. In
section \ref{sec2} we describe the multi-planetary system model  and give the basic equations used.
We  describe  the incorporation  of orbital circularization resulting  from  tides due to  the central star
and discuss the possible relevance of migration induced  through  interaction with the protoplanetary disk
when that was present,  though it is not included in our modelling.

We go on to develop a simple analytic model for a system in a three planet resonance 
undergoing circularization in section \ref{sec5}.
This is based on adopting a time averaged Hamiltonian for which contributions from
a maximum of four resonant angles are retained.  These are associated with the first order resonances
between the inner pair and outer pair of planets in the three planet system.
Dissipative effects due to orbital circularization are then added. These  cause  secular evolution
of the period ratios of the system.

We consider the case when  all four angles undergo small amplitude libration in section \ref{Lapmodel}.
In this case we obtain a solution for which the generalised three body  Laplace relation  is maintained throughout the evolution 
while the period ratios increasingly depart from strict commensurability.
We then go on to consider the case when only three of the resonance angles  undergo small amplitude
libration in section \ref{threeangev}. In this case the evolution is qualitatively similar.
However, the generalised three body Laplace relation is not maintained,  being replaced by a different relation 
between the period ratios which depends on details of the configuration.

In order to confirm predictions of the analytic model related to how the evolution time scale
scales with planet masses and tidal dissipation we carry out numerical simulations
of a three planet system with masses in the super-earth regime and which starts out 
close to a three planet resonance on section \ref{Numsimsec}.
In order to provide a focus,  in most cases the initial orbital parameters were taken to be those announced for the Kepler 60 system
(Steffen et al. 2013)   together with  the assumption of zero eccentricities.
Thus we do not strictly model the actual system but rather consider it
for the purpose of demonstrating the application of the analytical
formulation.
  
Finally  we discuss our results  and their application to providing constraints
on tidal dissipation,  focusing again on the Kepler 60  system,   and   conclude  in section \ref{sec6}.

\section{Model and basic equations}\label{sec2}
We consider a system of $N=3$ planets moving under their mutual gravitational attraction and that due to the central star.
The equations of motion are:
\begin{equation}
{d^2 {\bf r}_i\over dt^2} = -{GM{\bf r}_i\over |{\bf r}_i|^3}
-\sum_{j=1\ne i}^N {Gm_j  \left({\bf r}_i-{\bf r}_j \right) \over |{\bf
    r}_i-{\bf r}_j |^3} -{\bf \Gamma} +{\bf \Gamma}_{i} +{\bf \Gamma}_{r} \; ,
\label{emot}
\end{equation}

\noindent  where $M$, $m_i,$ $m_j,$ ${\bf r}_i$ and ${\bf r}_j$ denote the mass of
the central star, the mass of planet~$i,$ the mass of planet~$j,$  the position vector of planet
$i$ and the position vector of planet
$j,$ respectively.  The acceleration of the coordinate system based on
the central star (indirect term) is:
\begin{equation}
{\bf \Gamma}= \sum_{j=1}^N {Gm_j{\bf r}_{j} \over |{\bf r}_{j}|^3},
\label{indt}
\end{equation}

\noindent Tidal interaction with  the central star
 is dealt with through the addition of a  frictional  damping force taking the form (see eg.
Papaloizou~\& Terquem~2010)
\begin{equation}
{\bf \Gamma}_{i} =
 - \frac{2}{|{\bf r}_i|^2 t_{e,i}} \left( \frac{d {\bf r}_i}{dt} \cdot
{\bf r}_i \right) {\bf r}_i 
\label{Gammai}
\end{equation}

\noindent where $t_{e,i}$  is  the time scale
over which  the eccentricity  of an isolated planet damps.
Thus it is  the orbital  circularization time (see below).

\noindent  Relativistic
effects are included through ${\bf \Gamma}_{r}$ ( see Papaloizou \&
Terquem 2001).
Note that  in the
formulation above, eccentricity damping causes radial velocity damping,
which results in energy loss at constant angular momentum.  As a
consequence it causes  both the semi--major axis and the eccentricity to be reduced.

\subsection{Orbital migration}\label{TypeImigration}\label{sec3}
It is likely that systems of close orbiting planets  were not formed in their present locations
but were formed further out and then migrated inwards while the protoplanetary disk was still present 
(see eg. Papaloizou \& Terquem  2006;  Nelson \& Kley 2012 for reviews of orbital migration).
 However,  the  rate and    extent of  such migration is 
unclear  mainly because of uncertainties regarding the 
effectiveness of coorbital torques (e.g., Paardekooper \& Melema 2006 ) which can be sensitive to
the detailed local disk structure.  We note that  convergent  disk  migration leads naturally to  multiple systems
in resonant chains of the type considered here (eg. Cresswell \& Nelson 2006;  Papaloizou \& Terquem 2010).
Note that if they start out close to resonance,  these  configurations may be produced with  the system as a whole undergoing   little net radial migration.

As we are mainly  interested in the post formation evolution of the system,  we shall assume that 
although migration may have played a role in 
evolving the system into a  configuration that is close to a  commensurability,   during or just subsequent to 
planet formation,  the
protoplanetary disk has dispersed.   This has  the consequence  that  migration torques,  orbital circularization and changes to the orbital inclination
 arising  through disk-planet interaction are then absent.
Dissipative effects are assumed to  arise only through orbital circularization occurring as a result of tidal interaction
of the planets with the central star.

\subsection {Orbital circularization due to tides from the central star}\label{sec4}
The circularization timescale due to tidal interaction with the star
was obtained from   Goldreich~\& Soter (1966) in the form
\begin{equation}
t_{e,i}^s = \frac{4 m_i a_{i}^{13/2} Q'}{63G^{1/2}M^{3/2}R_{pi}^5} 
\label{teccs}
\end{equation}
\noindent where $m_i$ and  $a_i$ are the mass and  the semi--major axis of planet~$i$ respectively. 
Here $M$ is the mass of the central star and $R_{pi}$ is the radius of planet $i.$  The quantity
$Q'= 3Q/(2k_2),$ where $Q$ is the tidal dissipation function and $k_2$
is the Love number.  
 For solar system planets in the terrestrial mass
range, Goldreich \& Soter (1966) give estimates for $Q$ in the range
10--500 and $k_2 \sim 0.3$, which correspond to $Q'$ in the range
50--2500.  However,  
Ojakangas \& Stevenson (1986) argue that,  in the context of their episodic tidal heating model
for Io,  $Q$ can attain values of around unity at 
the solidus temperature.
 This is because  the  response
to tidal forcing  becomes  more
like that of a fluid with high viscosity than an elastic solid.
The rate of dissipation is  accordingly larger . The value of $Q$ is expected
to be a function of tidal forcing frequency as well as temperature that can 
attain a minimum value $\sim 1$ 
(see Ojakangas  \& Stevenson  1986 and  references therein). 
Although  this parameter must be 
regarded as being very uncertain for extrasolar planets, we should accordingly consider that
$Q$ may be of order unity under early post formation conditions where they may be near to  the solidus temperature.


We remark that, in our formulation, tidal interaction
with the star does not change the angular momentum of a single orbit but
causes it's energy to decrease. The physical basis for this is that the planets
rapidly attain pseudo-synchronization  (e.g., Ivanov \& Papaloizou 2007),   after which
they cannot store significant angular momentum  changes  through modifying
their  intrinsic  angular momenta.
For  the low mass planets considered here,  we  neglect tides induced on the central star (see eg. Barnes et al. 2009; Papaloizou \& Terquem 2010).
Thus  the orbital angular momentum 
and inclination  are not changed by the application of  ${\bf{\Gamma}}_{i}$ in equation 
(\ref{emot}).   




\section{Semi--analytic model for a system in a three planet resonance 
undergoing circularization}\label{sec5}

We  develop a semi--analytic model that shows how a  three
planet system undergoes  orbital evolution   driven by tidal circularization.
The coupling between the planets occurs because resonant angles are assumed to  librate.
But note that  there may be significant
deviations from exact commensurability.  
We begin by formulating equations governing  the system
without dissipation,  which is Hamiltonian,    we then  go on to add  in effects arising from orbital
circularization.

\subsection{Coordinate system}
We adopt Jacobi coordinates (eg. Sinclair~1975 ) for which the radius vector
 of  planet $i,$  ${\bf r }_i,$ is measured relative
to the  centre of mass of  the system comprised of $M$ and  all other  planets
interior to  $i,$ for  $i=1,2, 3.$ Here  $i=1$ corresponds to the
innermost planet and $i=3$ to the outermost planet.
 
 \subsection {Hamiltonian for the system without dissipation}\label{Hamilsec}
The  Hamiltonian,  correct to second order in the planetary masses,  can be written for the system without dissipation, 
in the form:
\begin{eqnarray} H & = &  \sum_{i=1}^3 \left({1\over 2}  m_i | \dot {\bf r}_i|^2
- {GM_{i}m_i\over  | {\bf r}_i|} \right)   \nonumber \\
& - &\sum_{i=1}^3\sum_{j=i+1}^3Gm_{i}m_j
\left({1 \over  | {\bf r}_{ij}|}  -  { {\bf r}_i\cdot {\bf r}_j
\over  | {\bf r}_{j}|^3}\right).
\end{eqnarray}
Here $M_{i}=M+m_i $ and
$ {\bf r}_{ij}= {\bf r}_{i}- {\bf r}_{j}.$

Assuming,  the planetary system is strictly coplanar, the equations governing the  motion  
within a fixed plane, about a dominant central mass,
 may be written in the form
(see, e.g., Papaloizou~2003; Papaloizou \& Szuszkiewicz~2005):

\begin{eqnarray}
\dot E_i &=& -n_i\frac{\partial H}{\partial \lambda_i}\label{eqnmo1}\\
\dot L_i &=& -\left(\frac{\partial H}{\partial \lambda_i}+\frac{\partial H}{\partial \varpi_i}\right)\\
\dot \lambda_i &=& \frac{\partial H}{\partial L_i} + n_i \frac{\partial H}{\partial E_i}\\
\dot \varpi_i &=& \frac{\partial H}{\partial L_i}.\label{eqnmo4}
\end{eqnarray}

 Here  and in what  follows until stated otherwise, 
 $m_i$ is replaced by he reduced mass  so that $m_i  \rightarrow m_{i}M/(M+m_{i}).$ 
The orbital  angular momentum of  planet  $i$ 
is $L_i$ and the
orbital  energy is $E_i.$
 The  mean longitude of planet $i$ is $\lambda_i = n_i (t-t_{0i}) +\varpi_i ,$
 with  $n_i  = \sqrt{GM_{i}/a_i^3}= 2\pi/P_i $ being  the  mean motion,  and
$t_{0i}$ denoting the time of periastron passage.  The semi-major axis and orbital period   
 of planet  $i$ are  $a_i$  and $P_i$  the respectively .
 The longitude of periastron is $\varpi_i.$
 The quantities $\lambda_i,$ $\varpi_i,$ $L_i$ and $E_i$ can be used to describe the dynamical
 system described above.

\noindent However,  we note that for motion around a central point  mass $M$ we have:
\begin{eqnarray}
     L_i &=&  m_{i}\sqrt{GM_{i}a_i(1-e_i^2)}, \\
     E_i &=& -{{GM_{i}m_{i}}\over{2a_i}},
\end{eqnarray}
where $M_{i} = M+m_{i},$  and $e_i$  the eccentricity of planet $i.$ 
Thus  by a simple transformation,   we  alternatively adopt   $\lambda_i,$ $\varpi_i,$ $a_i$ or equivalently $n_i,$  and $e_i$ as dynamical variables. 
  We comment that the difference between  taking $m_i$
to be the reduced mass rather than  the actual mass  of planet $i$ when evaluating $M_i$  in the  expressions for $L_i$ and $E_i$
is third order in the typical planet to star mass ratio and thus it may be neglected in the analysis below.


\subsection{Averaged Hamiltonian}\label{Havesec}
\noindent The Hamiltonian may quite generally
 be expanded in a Fourier series
involving linear combinations of the  five angular differences
$\varpi_1 -\varpi_2,$ $\varpi_2 -\varpi_3$ and
$\lambda_i - \varpi_i,  i=1,2,3. $ 

 Here we are interested
in the effects of  first order $p+1:p$  and $q+1:q$  commensurabilities
associated with the outer and inner pairs of planets respectively.
In this situation,  we expect that any of the four angles
$\Phi_1 = (q+1)\lambda_2-q\lambda_1-\varpi_1, $ 
$\Phi_2 = (q+1)\lambda_2-q\lambda_1-\varpi_2,$
$\Phi_3 = (p+1)\lambda_3-p\lambda_2-\varpi_2 $ and  
$\Phi_4 = (p+1)\lambda_3-p\lambda_2-\varpi_3$
may  be slowly varying.
 Following  standard practice
 (see, e.g.,  Papaloizou \& Szuszkiewicz~2005; Papaloizou \& Terquem~2010),
high frequency terms in the Hamiltonian are averaged out.
In this way,  only terms in the Fourier expansion involving  linear
combinations of $\Phi_1,$  $\Phi_2,$ $\Phi_3$ and $\Phi_4$
as argument are  retained. 

Working in the limit of small eccentricities that is applicable here,
 terms that are higher order than first  in the eccentricities can also be  discarded.
The  Hamiltonian   may  then be written
in the form:
\begin{equation} H=E_1+E_2+ E_3 +  H _{12} +H_{23},\label{Hamil0} \end{equation}
 where:
\begin{equation} H _{ij}= -\frac{Gm_im_j}{a_j}\left[  e_j C_{i,j}\cos (\Phi_{i+j-1})
- e_iD_{i,j}\cos (\Phi_{2i-1}) \right], \label{Hamil} \end{equation}
with:
\begin{eqnarray} C _{i,j} & = & {1 \over 2}\left(   x_{i,j}{db^{k}_{1/2}(x)\over dx} \Biggl|_{x=x_{i,j}}\Biggr. +(2k+1)b^{k}_{1/2}(x_{i,j})
-(2k+2)x_{i,j}\delta_{k,1} \right),   \label{Hamil1} \\
D _{i,j} & =& {1 \over 2}\left(   x_{i,j}{db^{k+1}_{1/2}(x)\over dx}\Biggl|_{x=x_{i,j}}\Biggr. +2(k+1)b^{k+1}_{1/2}(x_{i,j})  \right) . \label{Hamil2} 
\end{eqnarray}
Here the integer $k=q$ for the inner pair $(i=1,j=2)$ and $k=p$ for the outer pair
$(i=2,j=3)$  with  $b^{k}_{1/2}(x)$ denoting the usual Laplace coefficient
(e.g., Brouwer \& Clemence 1961;  Murray and Dermott 1999)
with the argument $x_{i,j} = a_i/a_j.$

\subsection{Incorporation of orbital circularization due to tidal interaction with the central star}
 
 Using equations~(\ref{eqnmo1})--(\ref{eqnmo4})
together with equation~(\ref{Hamil0}) we may first obtain the equations of motion
without the effect of circularization due to tidal interaction with the central star. 
Having obtained  the forme, the effect of the latter may be added in 
(see eg. Papaloizou~2003). Following this procedure, we obtain:
\begin{align}
\frac{d e_1}{dt}&=
\frac{m_2a_1 n_1D_{12}}{a_2 M_1}\sin\Phi_1 - \frac{e_1}{t_{e,1}^s}, \label{eqne1}\\
\!\!\!\!\!\!\!\!\!\!\!
\frac{d e_2}{dt}&=
-\frac{n_2 }{M_2}
\left(m_1C_{12}\sin\Phi_2-\frac{a_2}{a_3}m_3D_{23}\sin\Phi_3\right)
-\frac{e_2}{t_{e,2}^s},
\label{eqne2}\\
\!\!\!\!\!\!\!\!\!\!
\frac{d e_3}{dt} &=
-\frac{m_2 n_3C_{23}}{ M_3}\sin\Phi_4 - \frac{e_3}{t_{e,3}^s}, \hspace{1cm}{\rm } \label{eqne3}\\
\!\!\!\!\!\!\!\!\!\!
\dot n_1 &=
-\frac{3qn_1^2m_2a_1}{M_1a_2}\left(C_{12}e_2\sin\Phi_2-D_{12}e_1\sin\Phi_1 \right)
+\frac{3n_1e_1^2}{t_{e,1}^s}, \label{eqntid0}\\
\!\!\!\!\!\!\!\!\!\!
 \dot n_2 &=
\frac{3(q+1)n_2^2m_1}{M_2}\left(C_{12}e_2\sin\Phi_2-D_{12}e_1\sin\Phi_1 \right)\nonumber\\
\!\!\!\!\!\!\!\!\!\!
   &-
\frac{3pn_2^2m_3a_2}{M_2a_3}\left(C_{23}e_3\sin\Phi_4-D_{23}e_2\sin\Phi_3 \right)
+\frac{3n_2e_2^2}{t_{e,2}^s}, \label{eqntid} \\
\dot n_3 &=
\frac{3(p+1)n_3^2m_2}{M_3}\left(C_{23}e_3\sin\Phi_4-D_{23}e_2\sin\Phi_3 \right)
+\frac{3n_3e_3^2}{t_{e,3}^s},  \label{eqntid1}\\
\!\!\!\!\!\!\!\!\!\!
\dot \Phi_1 &=
(q+1)n_2-q n_1+\frac{n_1}{e_1}\frac{m_2 a_1}{M_1 a_2} D_{12}\cos \Phi_{1},
 \label{eqntid20}\\
\!\!\!\!\!\!\!\!\!\!
\dot \Phi_2&= 
(q+1)n_2-q n_1-\frac{n_2}{e_2}\left[ \frac{m_1}{M_2} C_{12}\cos \Phi_{2}-\frac{m_3a_2}{M_2 a_3}D_{23}\cos \Phi_{3}\right]  ,
 \label{eqntid3}\\
\!\!\!\!\!\!\!\!\!\!
\dot \Phi_3 &= 
(p+1)n_3-p n_2-\frac{n_2}{e_2}\left[ \frac{m_1}{M_2} C_{12}\cos \Phi_{2}-\frac{m_3a_2}{M_2 a_3}D_{23}\cos \Phi_{3}\right]  ,
 \label{eqntid4}\\
 \!\!\!\!\!\!\!\!\!\!
\dot \Phi_4 &=
(p+1) n_3-pn_2-\frac{n_3}{e_3}\frac{m_2}{M_3 }C_{23}\cos \Phi_{4}  .
 \label{eqntid5}\\
\nonumber
\end{align}







\noindent  At this point we note that in the analysis below we 
use equations    (\ref{eqne1}) - (\ref {eqntid5}) to
calculate  perturbations to the  orbital elements and resonant angles
correct to first order in the typical planet to central star has ratio.
For this purpose from now on  we  adopt the actual mass  of planet $i$ for $m_{i}, $ rather than the reduced mass and replace $M_i$ by $M,$
as any associated  corrections  will be second order.  Similarly the difference between using 
 actual or reduced masses when evaluating the circularization  times may be neglected as any corrections to those will also be correspondingly small.
Accordingly $m_i$ will be identified as being the actual mass of planet $i$ everywhere  from now on.

\noindent The terms involving  the circularization times
$t_{e,i}^s$  are associated with effects  arising from  the tides
 raised by the central mass on  planet $i.$
Notably, the terms $\propto e_i^2/t_{e,i}^s$
in equations~(\ref{eqntid0}), (\ref{eqntid}) and (\ref{eqntid1})
arise on account of  the orbital energy dissipation occurring as a result of circularization
at the lowest order in $e_i,$ and the disturbing masses, for which this  appears.
We shall assume throughout that such terms, through being retained, can be  of lower order than those proportional to the disturbing masses, $m_i.$
However, we shall assume that expressions that are of second or higher order in the ratio of these quantities  may be neglected.


\subsubsection{Energy and angular momentum conservation}

In the absence of circularizing tides $(t_{e,i}^s\rightarrow \infty),$   the total energy $E\equiv H$ and angular momentum
$L=L_1+L_2+L_3$ are conserved.
When circularizing tides act, the total angular momentum is conserved but energy is lost according to:
\begin{equation}
\frac{dH}{dt}= \sum_{i=1}^3\frac{2e_i^2E_i}{t_{e,i}^s},
\label{Econ}
\end{equation}
  Equation (\ref{Econ}) follows from the averaged equations, 
with $H$ being the unperturbed Hamiltonian,  provided the relative commensurabilites are assumed to be satisfied
to within the order of the perturbing masses and    relative corrections of  the order of the  of the squares of the perturbing masses
are neglected.   

\subsection{The case when four angles undergo small amplitude libration}\label{Lapmodel}
\noindent In practice we find that a  quasi-stationary solution is  possible for which $\Phi_1$ and $\Phi_3$
librate about angles close to  zero
and  $\Phi_2$ and $\Phi_4$ librate about angles close to $\pi.$ We describe  this in this section.
We begin by relating the resonance angles to the eccentricities.

\subsubsection{Specification of the resonance angles in terms of the eccentricities 
}
We assume that the system  governed by (\ref{eqne1}) - (\ref{eqntid5}) 
 undergoes very slow secular evolution on a time scale very much longer than either  the 
characteristic circularization time, or the characteristic  time scales  associated with perturbations  being of the order of
$M(n_im_i)^{-1}.$ We suppose  that under these conditions 
a quasi-steady state exists, for which the time derivatives of the eccentricities may be neglected.  From  (\ref{eqne1}) we obtain 

\begin{equation} e_1     = 
\frac{m_2a_1 n_1D_{12}t_{e,1}^s}{a_2 M}\sin\Phi_1\equiv \gamma_1\sin\Phi_1 .\label{e1ph}\end{equation}  
Similarly from  (\ref{eqne3}) we obtain 
\begin{equation} e_3     = 
-\frac{m_2 n_3C_{23}t_{e,3}^s}{ M}\sin\Phi_4 \equiv \gamma_3\sin\Phi_4\end{equation}  
and from (\ref{eqne2}) we obtain
\begin{equation} e_2     = 
\frac{ n_2t_{e,2}^s}{ M}\left(\frac{m_3a_2 D_{23}}{a_3}\sin\Phi_3    -  m_1 C_{12}\sin\Phi_2 \right)\equiv
 -\gamma_{22} \sin\Phi_2+\gamma_{23}\sin\Phi_3.\label{eqne31} \end{equation}
    
 For the case where  all the resonant  angles are stationary or undergo small amplitude librations,   equations  (\ref{eqntid3})  and  (\ref{eqntid4}) give the
 generalized  three body Laplace resonance condition 
$(p+q+1)n_2-qn_1-(p+1)n_3=0.$
Using equations   (\ref{eqntid0}) - (\ref{eqntid1})  we find that  the time derivative
of  this condition implies that
\begin{eqnarray} 
&&\hspace{-6mm}\beta_{1}e_1 \sin\Phi_1+\beta_{22}e_2\sin\Phi_2
 +\beta_{23} e_2\sin\Phi_3+\beta_{3}e_3\sin\Phi_4   \frac{ }{}\nonumber \\
 &&\hspace{-6mm}= \frac{ n_1 q e_1^2}{ t_{e,1}^s}  
 +\frac{ n_3 (p+1) e_3^2}{ t_{e,3}^s} - \frac{ n_2 (q+1+p) e_2^2}{ t_{e,2}^s},\label{e2ph} 
 \end{eqnarray}  
where
\begin{eqnarray} 
\beta_{1}& =&-D_{12}\left(\frac{(q+1)(p+q+1)n_2^2m_1}{M}   +\frac{q^2n_1^2m_2a_1}{a_2M}\right)    , \hspace{2mm} \beta_{22}=-\frac{\beta_1C_{12}}{D_{12}}\nonumber \\
 \beta_{23}& =&D_{23}\left(\frac{p(p+q+1)n_2^2m_3a_2}{a_3M}   +\frac{(p+1)^2n_3^2m_2}{M}\right)    ,\hspace{2mm} {\rm and} \hspace{2mm} \beta_{3}=-\frac{\beta_{23}C_{23}}{D_{23}}.\label{betadef}
 \end{eqnarray}  
Now we use equations (\ref{e1ph})-(\ref{e2ph}) to eliminate the angles  $\Phi_i$
in favour of the eccentricities $e_i.$ After having done this we determine the
rate of the change of the separation of the system from precise commmensurability. 
To do this we begin by using
 equations (\ref{eqntid0}) and (\ref{eqntid})  to obtain
\begin{eqnarray} 
&& \hspace{-6mm}\frac{d}{dt} \left((q+1)n_2-qn_1\right)=\nonumber\\
&&\hspace{-6mm} \alpha_{1} e_1\sin\Phi_1+\alpha_{22}e_2\sin\Phi_2
 +\alpha_{23} e_2\sin\Phi_3+e_3\alpha_{3}\sin\Phi_4+
 \frac{3n_2(q+1) e_2^2}{ t_{e,2}^s} - \frac{ 3q n_1 e_1^2}{ t_{e,1}^s}, \label{sepdot}
 \end{eqnarray}  
where
\begin{eqnarray} 
\alpha_{1}& =&-3D_{12}\left(\frac{(q+1)^2n_2^2m_1}{M}   +\frac{q^2n_1^2m_2a_1}{a_2M}\right)    , \hspace{2mm} \alpha_{22}=-\frac{\alpha_1C_{12}}{D_{12}}\nonumber \\
 \alpha_{23}& =&3D_{23}\frac{p(q+1)n_2^2m_3a_2}{a_3M}  ,\hspace{2mm} {\rm and} \hspace{2mm} \alpha_{3}=-\frac{\alpha_{23}C_{23}}{D_{23}}. \label{alphadef}
 \end{eqnarray}  

\noindent We can now use the specification of the resonant angles in terms
of the ecentricities given through equations (\ref{e1ph})-(\ref{e2ph}) 
in equation (\ref{sepdot}) which will then express  the rate of change of the mean motion separation
from resonance of planets $1$ and $2$ in terms of the eccentricities and semi-major axes  in the form: 
%
\begin{eqnarray}
&& \hspace{-6mm}\frac{d}{dt} \left((q+1)n_2-qn_1\right)=  -A_1\frac{ n_1  e_1^2}{ t_{e,1}^s}  -A_2\frac{ n_2  e_1^2}{ t_{e,2}^s} 
 -A_3\frac{ n_3  e_3^2}{ t_{e,3}^s}\label{Delev} 
 \end{eqnarray}
where
\begin{eqnarray}
 &&\hspace{-6mm}A_1= - \frac{ \alpha_1t_{e,1}^s }{n_1\gamma_1}+3q+ \left( \frac{ \beta_1t^s_{e,1} }{n_1\gamma_1}-q\right)
\frac{(\alpha_{23}\gamma_{22} +\alpha_{22}\gamma_{23})}{ (\beta_{22}\gamma_{23} +\beta_{23}\gamma_{22})}
\end{eqnarray}
\begin{eqnarray}
 &&\hspace{-6mm} A_2 = -3(q+1)- \frac{ t_{e,2}^s }{n_2}
\frac{(\alpha_{23}\beta_{22} -\alpha_{22}\beta_{23})}{ (\beta_{22}\gamma_{23} +\beta_{23}\gamma_{22})}
+(p+q+1)\frac{(\alpha_{23}\gamma_{22} +\alpha_{22}\gamma_{23})}{ (\beta_{22}\gamma_{23} +\beta_{23}\gamma_{22})}\
\end{eqnarray}
\begin{eqnarray}
 &&\hspace{-6mm} A_3 =- \frac{ \alpha_3t_{e,3}^s }{n_3\gamma_3}+\left( \frac{ \beta_3t^s_{e,3} }{n_3\gamma_3}-(p+1)\right)
 \frac{(\alpha_{23}\gamma_{22} +\alpha_{22}\gamma_{23})}{ (\beta_{22}\gamma_{23} +\beta_{23}\gamma_{22})}
\end{eqnarray}
Note that we can  use the definitions of  the coefficients $\alpha_i, \alpha_{ij}, \beta_i, \beta_{ij}, \gamma_i, \gamma_{ij}$
given by (\ref{alphadef}), (\ref{betadef}) and    (\ref{e1ph}) - (\ref{eqne31}) 
 to substitute for them,  in terms 
of the semi-major axes and the circulazisation times,  in  equation (\ref{Delev} ).
\subsubsection{Specification of the eccentricities in terms of the deviation from commensurability}
\noindent In order to complete the calculation of the orbital evolution away from commmmensurability,
we need to specify the eccentricities as a function of the semi-major axes.
To do this we use equations (\ref{eqntid20})- (\ref{eqntid5})
 together with the assumption  that  the angles $\Phi_i$  are close to their equilibrium values.
   Later we shall present numerical simulations for which   $\Phi_1$ and $\Phi_3$
 are close to zero
and  $\Phi_2$ and $\Phi_4$ are close to  $\pi.$ Accordingly  we  adopt those values for the $\Phi_i$
while setting  their  time derivatives  to be zero.  Equations (\ref{eqntid20}) - (\ref{eqntid5}) 
then specify the eccentricities through
\begin{align}
e_1 &=
-\frac{n_1}{((q+1)n_2-q n_1)}\frac{m_2 a_1}{M a_2} D_{12},
 \label{eqntid40}\\
\!\!\!\!\!\!\!\!\!\!
e_2&= 
-\frac{n_2}{((q+1)n_2-q n_1)}\left( \frac{m_1}{M} C_{12}+\frac{m_3a_2}{M a_3}D_{23}\right)  ,
 \label{eqntid43}\\
 \!\!\!\!\!\!\!\!\!\!
e_3 &=
-\frac{n_3}{((p+1) n_3-pn_2)}\frac{m_2}{M }C_{23} .
 \label{eqntid45}\\
\nonumber
\end{align}

\noindent When all four angles librate, 
we have the generalised three planet  Laplace resonance condition that
 
\noindent $(p+1)n_3 - pn_2=(q+1)n_2 - qn_1$~(see~above).  
This   state applies to systems slightly more widely spaced than would be the case
for precise commensurability, so that

\noindent  $(p+1)n_3 - pn_2~=~(q+1)n_2~-~qn_1~\equiv~\Delta_{21}$ is negative, the  period ratios
exceed  the commensurable values  and
the eccentricities will be  positive.
 
We remark that for the alternative librating state for which  $\Phi_1$ and $\Phi_3$
are close to $\pi,$
and $\Phi_2$ and $\Phi_4$ are close to zero, equations (\ref{eqntid40}) -(\ref{eqntid45}) hold with a reversal of sign.
Thus that situation applies when $(p+1)n_3-pn_2~=~(q+1)n_2-qn_1~\equiv~\Delta_{21}$ is positive,  so that
the period ratios are smaller than the commensurable values, ensuring again that  the eccentricities will be positive.
Bearing in mind this proviso, as it depends only on the squares
of the eccentricities obtained from (\ref{eqntid40}) -(\ref{eqntid45}),
the anlysis presented below can applied to both librational states.

We are now able to use equation (\ref{Delev}) together with  equations  (\ref{eqntid40}) - (\ref{eqntid45}) 
to obtain an equation governing the evolution of $\Delta_{21}$  in the form: 
\begin{eqnarray}
&& \hspace{-6mm}\Delta_{21}^2\frac{d\Delta_{21}}{dt}   =  
- \frac{A_1 n_1^3}{ t_{e,1}^s}\left(\frac{m_2 a_1 D_{12}}{M a_2}\right)^2
  -\frac{ A_2n_2^3 }{ t_{e,2}^s}\left( \frac{m_1}{M} C_{12}+\frac{m_3a_2}{M a_3}D_{23}\right)^2 
 -\frac{A_3 n_3^3}{ t_{e,3}^s}\left(\frac{m_2 C_{23}}{M}\right)^2,\nonumber \\
&& \hspace{-6mm}{\rm  }\label{Delevfin} 
 \end{eqnarray}
 where after eliminating the $\alpha_i, \alpha_{ij}, \beta_i, \beta_{ij}, \gamma_i, \gamma_{ij}$
 using  (\ref{alphadef}), (\ref{betadef}) and    (\ref{e1ph}) - (\ref{eqne31}) as indicated above, we may write
\begin{eqnarray}
 &&\hspace{-6mm}A_1= 3(q+1)\left((q+1)\frac{n_2^2m_1a_2}{n_1^2m_2a_1}+q\right) - 
(q+1){\cal A}\left((p+q+1)\frac{n_2^2m_1a_2}{n_1^2m_2a_1} +q\right),
\end{eqnarray}
\begin{eqnarray}
 &&\hspace{-6mm} A_2 = -3(q+1)-{\cal B}
+(p+q+1){\cal A},
\end{eqnarray}
\begin{eqnarray}
 &&\hspace{-6mm} A_3 =- 3p(q+1)\frac{n_2^2m_3a_2}{n_3^2m_2a_3}+
p{\cal A}\left( (p+q+1)\frac{n_2^2m_3a_2}{n_3^2m_2a_3}+p+1\right), \hspace{3mm}{\rm with}
\end{eqnarray}

\begin{eqnarray}
&&\hspace{-6mm}
{\cal B} =
-\frac{3Mm_2{\cal B}_1}{n_2^2a_2{\cal D}},    \hspace{3mm}{\rm and}\\
&&\hspace{-6mm}
{\cal A} =\frac{{\cal A}_1}{{\cal D}},  \hspace{3mm}{\rm with}\\
&&\hspace{-6mm}{\cal B}_1=   p^2q^2m_3Ma_1 a_2n_1^2n_2^2
 +(p+1)^2q^2m_2Ma_1 a_3n_1^2n_3^2
+(p+1)^2(q+1)^2m_1Ma_2 a_3n_2^2n_3^2,
 \\
&&\hspace{-6mm}{\cal A}_1 = 3M^2\left(  q^2m_2m_3a_1 n_1^2
  + (q+1)(q+1+p)m_1m_3a_2 n_2^2\right),    \hspace{3mm}{\rm and}\\
&&\hspace{-6mm}{\cal D} = (p+q+1)^2m_1m_3M^2a_2 n_2^2   + (p+1)^2m_1m_2M^2a_3 n_3^2
 +q^2m_2m_3M^2a_1 n_1^2.
\end{eqnarray}
\subsection{Time dependent evolution of the separation from resonance}
Equation (\ref{Delevfin}) governs the time dependent evolution of $\Delta_{21}=(q+1)n_2-qn_1 = (p+1)n_3-pn_2$ as a function of time.
We first remark that each of the three contributions to the right hand side of  (\ref{Delevfin})
are readily seen to be non negative  and so $\Delta_{21}$ decreases monotonically with time.
 When $\Delta_{21} < 0,$ this means that $|\Delta_{21}|$ will subsequently  increase with time,
 corresponding to an increasing departure from strict resonanance.
The expression of the  conservation of the total angular momentum
of the system can be used to enable each of the $n_i$ to be expressed in terms of $\Delta_{21}$ 
which may then be found by integration. However, when the system still 
remains close to resonance, the situation may be simplified
  by noting that
 the relative period separation from resonance may increase significantly while the mean motions themselves
change negligibly. 
  Under these conditions  the right hand side of (\ref{Delevfin}) may be approximated as being  constant.
To simplify notation we define a quantity ${\cal T}$  by writing this equation as
\begin{eqnarray}
  \Delta_{21}^2\frac{d\Delta_{21}}{dt}  = -n_1^3/(3{\cal T}). \label{Delevfin1}
\end{eqnarray}
Under the assumption of a constant right hand side, the solution is given by
\begin{eqnarray}
&&\hspace{-6mm}
\frac{ \Delta_{21}}{n_1} = -\left(\frac{(t-t_0)}{{\cal T}}\right)^{1/3}, \label{Anfit}
\end{eqnarray}
where $t_0$ is a constant of integration. This is such that at a putative time $t=t_0,$ $\Delta_{21}=0.$
$|\Delta_{21}|$ is then monotonically increasing for $t > t_0.$ 
We remark that   corresponding tendency for the departure of the relative period ratio  from strict commensurability
to increase $\propto t^{1/3},$ for systems comprising two planets close to resonance, has been noted by Papaloiozu (2011),
Lithwick \& Wu (2012) and Batygin \& Morbidelli (2012).

 
\subsection{The case when three  angles undergo small amplitude libration}\label{threeangev}
\noindent We have considered quasi-stationary solutions for which the  angles  $\Phi_1$ and $\Phi_3$
 librate about   angles close to zero
and  $\Phi_2$ and $\Phi_4$ librate about angles close to  $\pi.$ In this case the generalised  three body Laplace resonance condition 
$(p+1) n_3-pn_2 = (q+1) n_2-qn_1$ applies.  However, we can also consider the possibility that this relation is relaxed
with only three of the angles in a state of libration, while the fourth circulates.
This  situation  is of interest as it was found by Papaloizou \& Terquem (2010) to occur
 in simulations of the tidal evolution of the  HD40307 system
away from a 4:2:1 resonance $(p=q=1).$

There are two possible choices for the circulating angle that allow secular evolution without the generalised three body  Laplace  resonance  condition  holding,  namely $\Phi_2$ or  $\Phi_3.$
We shall consider both these cases below assuming that liberating angles remain close to the same values as in the previous case,
though it is relatively simple to generalize the discussion to other examples.
 We proceed by closely following  the analysis of the previous section,
noting that terms involving the circulating angle are omitted as this is assumed to make no contribution to
the time averaged dynamics. Thus equations (\ref{e1ph})-(\ref{eqne31}) hold but with the contribution from the
circulating angle set to zero. They then suffice to determine the three remaining angles in terms of the eccentricities
without having to add the constraint given by (\ref{e2ph}) which expresses the fact that the generalized three body
Laplace resonance condition holds for all time.

However equation  (\ref{sepdot}) for $d((q+1) n_2-qn_1)/dt$ still holds but again with the contribution from the circulating
angle omitted. Using this with the modified forms of (\ref{e1ph})-(\ref{eqne31})  described above being  used to 
eliminate the librating angles in favour of the eccentricities, we obtain 
\begin{eqnarray}
&& \hspace{-6mm}\frac{d}{dt} \left((q+1)n_2-qn_1\right)=  -B_1\frac{ n_1  e_1^2}{ t_{e,1}^s}  -B_2\frac{ n_2  e_2^2}{ t_{e,2}^s} 
 -B_3\frac{ n_3  e_3^2}{ t_{e,3}^s},\label{Del21ev} 
 \end{eqnarray}
where
\begin{eqnarray}
 &&\hspace{-6mm}B_1= 
3(q+1)^2\frac{n_2^2m_1a_2}{n_1^2m_2a_1} +3q(q+1),
\end{eqnarray}
\begin{eqnarray}
 &&\hspace{-6mm} B_2 = -3(q+1)+ {\cal E}_2  \hspace{4mm} {\rm and}
\end{eqnarray}
\begin{eqnarray}
 &&\hspace{-6mm} B_3 =
 -3p(q+1)\frac{n_2^2a_2m_3}{n_3^2a_3m_2}.
 \end{eqnarray}
Here, either  
\begin{eqnarray}
 &&\hspace{-6mm} {\cal E}_2 = \frac{3q^2n_1^2a_1m_2}{n_2^2 a_2m_1} + 3(q+1)^2 
\end{eqnarray}
if $\Phi_2$ librates,   or alternatively,    ${\cal E}_2=  
-3p(q+1)$
if $\Phi_3$ librates.  

In order to complete the determination of the evolution, as before we may use equations (\ref{eqntid20})- (\ref{eqntid5}) to specify the eccentricities in terms of the semi-major
axes. However, as above , we must omit the equation  derived from  setting the time derivative of the circulating angle to zero
as well as    contributions potentially  arising from the  circulating  angle in the remaining equations.
One then finds that  equations  (\ref{eqntid40}) and  (\ref{eqntid45}) can be taken over unaltered. On the other hand,  equation (\ref{eqntid43}) has to be  replaced by 
\begin{eqnarray}
e_2&=& 
-\frac{n_2}{ \Delta} F_2  ,
 \label{eqntid46}
\end{eqnarray}
where either  $F_2= m_1C_{12} /M$ with $\Delta = (q+1)n_2-qn_1=\Delta_{21},$  when $\Phi_2$ librates,  or $F_2= m_3a_2D_{23}/(M a_3)$
with $\Delta=(p+1)n_3-p n_2=\Delta_{32}$  when $\Phi_3$ librates.

\subsubsection{The relation between $\Delta_{32}$ and  $\Delta_{21}$}\label{Threeangsub}
As the generalized three body  Laplace relation is no longer satisfied,  we must find an alternative means of relating $\Delta_{32}=(p+1)n_3-pn_2$
 and $\Delta_{21}=(q+1)n_2-qn_1.$
This can be done by writing down the equation for the time derivative of $\Delta_{32}$ obtained from  
(\ref{eqntid0}) - (\ref {eqntid1}). Proceeding in the same manner
as for  $d\Delta_{21}/dt,$ , we can write this in terms of the semi-major axes and eccentricities in the form:


\begin{eqnarray}
&& \hspace{-6mm}  \frac{d\Delta_{32}}{dt} = -K_1\frac{ n_1  e_1^2}{ t_{e,1}^s}  -K_2\frac{ n_2  e_2^2}{ t_{e,2}^s} 
 -K_3\frac{ n_3  e_3^2}{ t_{e,3}^s},\label{Del32ev} 
 \end{eqnarray}
where
\begin{eqnarray}
 &&\hspace{-6mm}K_1= -\frac{ 3p(q+1) n_2^2a_2m_1}{n_1^2a_1m_2} ,
\end{eqnarray}
\begin{eqnarray}
 &&\hspace{-6mm} K_2 = 3p+S_2  \hspace{4mm} {\rm and}
\end{eqnarray}
\begin{eqnarray}
 &&\hspace{-6mm}K_3=  3p(p+1) +     \frac{ 3p^2 n_2^2a_2m_3}{n_3^2a_3m_2} .
\end{eqnarray}
Here, either  
\begin{eqnarray}
&&S_2= - 3p(q+1)
\end{eqnarray}
 if  $\Phi_2$ librates,  or  alternatively,
 \begin{eqnarray}
 && S_2=   \frac{ 3(p+1)^2 n_3^2a_3m_2}{n_2^2a_2m_3}+      3p^2 
  \end{eqnarray}
if $\Phi_3$ librates. 
 Equations (\ref{Del21ev}) and (\ref{Del32ev}) together with the conservation of angular momentum
determine the evolution of the system. This is qualitatively similar to the case where the strict  generalized three body Laplace resonance applies.
Here we focus on the relation between $\Delta_{32}$ and $\Delta_{21}$  which can be obtained by dividing equation
(\ref{Del32ev}) by equation (\ref{Del21ev}) . This gives 

\begin{eqnarray}
&& \hspace{-6mm}  \frac{d\Delta_{21}}{d \Delta_{32}} = \frac{B_1n_1  e_1^2/ t_{e,1}^s  +B_2 n_2  e_2^2/t_{e,2}^s 
 +B_3 n_3  e_3^2/ t_{e,3}^s} {K_1 n_1  e_1^2/t_{e,1}^s  +K_2n_2  e_2^2/t_{e,2}^s 
 +K_3 n_3  e_3^2/ t_{e,3}^s},\label{Delratev} 
 \end{eqnarray}

\noindent Equations  (\ref{eqntid40}),  (\ref{eqntid45}) and  (\ref{eqntid46}) can be used  to  specify
the eccentricities.  Equation (\ref{Delratev}) then determines  the ratio $\Delta_{32}/\Delta_{21}.$
This depends on the ratios of the circularization times for the different planets.
When the circularization time is proportional to a  power of the semi-major axis and the system is close to resonance,
such that the  ratios of the semi-major axes of the different planets can be taken to be fixed,
 we can approximate the ratio  as being constant.  As an example, we consider the case when $\Phi_3$ librates and $t_{e,3}\rightarrow \infty.$
This is the case that was found by Papaloizou \&  Terquem (2010)   to occur  in their simulations of the tidal evolution of the  HD40307 system.
It is also a reasonable  limit   to consider as  circularization  is  reasonably expected to be significantly less effective
for the outermost planet.  In this case we obtain

\begin{eqnarray}
&& \hspace{-6mm}  \frac{\Delta_{21}}{ \Delta_{32}} =\frac{{\cal D}_1}{{\cal D}_2},\hspace{6mm}  {\rm where}  \label{Delratevex}\\ 
&& \hspace{-6mm} {\cal D}_1= 
 \frac{(q+1)^2n_2^2  a_2m_1}{ n_1^2a_1m_2}      +q(q+1)- (q+1)(p+1)W, \hspace{6mm}  {\rm and}\\ 
 && \hspace{-6mm} {\cal D}_2= 
-\frac{p(q+1)n_2^2a_2m_1}{ n_1^2a_1m_2}  +  \left( p(p+1)+((p+1)^2n_3^2  a_3m_2)/(n_2^2a_2m_3)\right)W\\ 
&&\hspace{2cm}        
,\nonumber  
 \end{eqnarray}
where
\begin{eqnarray}
&& \hspace{-6mm}  W = \left(\frac{ m_3n_2a_2^2D_{23}}{ m_2n_1a_1a_3D_{12}}\right)^2\left( \frac{n_2t_{e,1}^s}{n_1t_{e,2}^s}\right)
\left(\frac{\Delta_{21}}{\Delta_{32}}\right)^2
 \end{eqnarray}
Equation (\ref{Delratevex}) is seen to provide a cubic equation for the ratio, $x=\Delta_{32}/\Delta_{21}.$
This is readily seen to have one positive real root that is appropriate for the assumptions we have made.
This has a particularly simple form  when the ratio of the circularization times, $t_{e,1}^s /t_{e,2}^s$ is assumed to be small.
Then as $x$ vanishes in the limit that this approaches zero, we see that denominator on the right hand side of (\ref{Delratevex}) 
must vanish in this limit. This gives an expression for $x$ through

\begin{eqnarray}
&& \hspace{-6mm}  x^2 = \left(\frac {p(p+1) +    (p+1)^2n_3^2  a_3m_2)/(n_2^2a_2m_3)}{p(q+1)}\right)
 \left( \frac{m_3^2n_2a_2^3D_{23}^2t_{e,1}^s}{m_1m_2n_1a_1a_3^2D_{12}^2t_{e,2}^s}\right)       
.\label{exroot} 
 \end{eqnarray}
 Thus when three angles librate as considered above,  when the circularization times scale as a power of the semi-major axes,  
$x$  can  also be approximately  constant during the evolution as was indicated by Papaloizou \& Terquem (2010). However, the magnitude of the constant  depends  
 on the ratios of the circularization times for  different planets.  It is accordingly uncertain,  though the above discussion
 shows that $x\rightarrow 0$ when the circularization time for the innermost planet is very much shorter than the
corresponding time for  the next innermost planet
 which is in turn very much shorter than the corresponding time for the outermost  planet.
 
  Once the value of $x$ has been determined, the evolution of the system can be obtained
 by solving (\ref{Del21ev}) in the same way  as for the case when  the strict generalised three body  Laplace resonance condition holds.
As in that case we find that the deviation from exact commensurability increases  $\propto t^{1/3}.$  
However, as we consider simulations of cases where the four angles ultimately librate  with the Laplace resonance holding,
below, we shall not consider the case of three angle libration in any further detail in this paper.

\subsubsection{Comparison with previous work}
Batygin \&  Morbidelli (2012) have considered the evolution of a three planet system under orbital circularization
towards a librating state. They considered systems for which three angles circulated and one librated.
Most of the analytic discussion  and all of the numerical work in this paper is concerned with systems where four angles librate.
However, systems for which only three angles librate were considered in sections \ref{threeangev} and \ref {Threeangsub} above.

Batygin \&  Morbidelli (2012) give equations governing the evolution of such a system once it has attained a librating state
although no analytic solutions were subsequently discussed. However, solutions of the kind 
considered in sections  \ref{threeangev} and \ref {Threeangsub} are readily obtained from their equations (47).
These give three equations for the quantities $m_1\sqrt{a_1},   m_2\sqrt{a_2}$ and $m_3\sqrt{a_3}.$
These can be  readily converted into two equations for  $\Delta_{21}$ and $ \Delta_{32}$ as defined above.
The ratio of these equations will then lead to the equivalent of equation  (\ref{Delratev}) above.

\section{Numerical Simulations}\label{Numsimsec}
In this section we give  the results of numerical simulations of thee planet
systems close to a three planet resonance under the influence of tidal circularization due to the central star.
We focus on systems with parameters with initial conditions near to, or corresponding to
those appropriate to the Kepler  60 system. 
Thus the initial conditions were taken to correspond to the tabulated orbital elements given by Steffen et al. (2013) with the added 
stipulation that 
the initial eccentricities are zero. In doing this  we remark  that the actual eccentricities  cannot be strictly zero with the consequence that  we model
a system that is close to rather than identical to the actual system. This should be adequate for illustrating the general theory and determining
characteristic evolutionary time scales.

\begin{figure}
\begin{center}
\epsfig{file=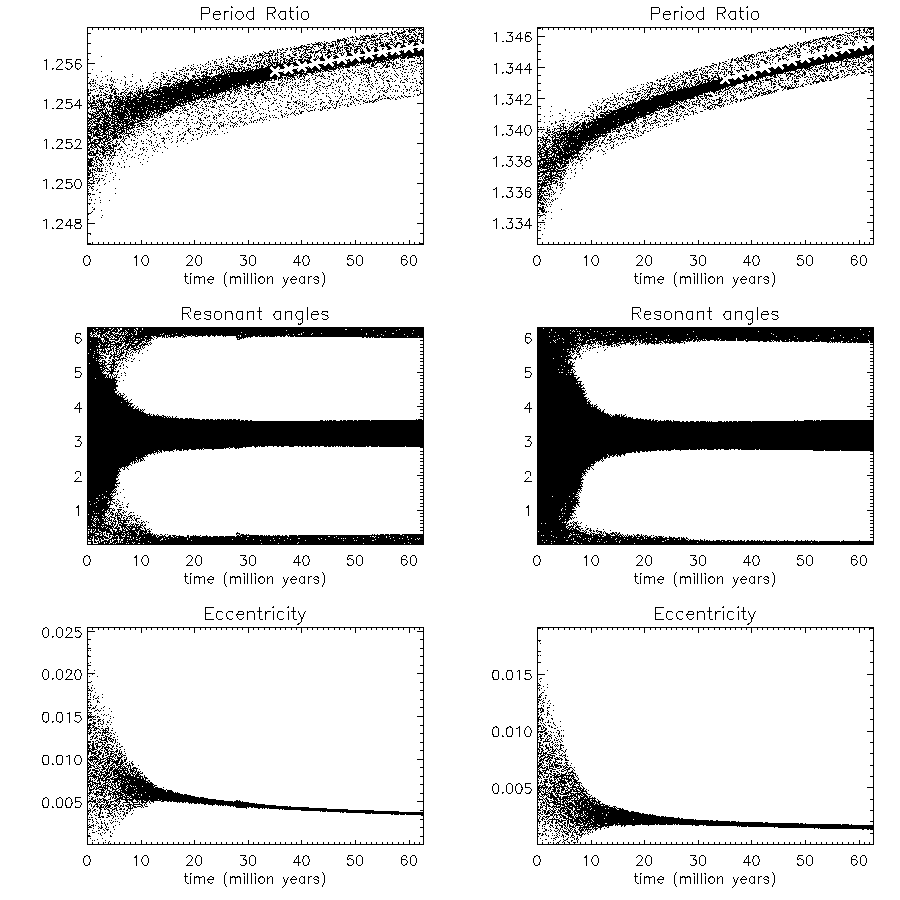
, width=16cm,height=18cm} 
 \caption{ Results for $Q'~=~2.5({\overline \rho}/2.4)^{-5/3}. $ 
The upper left panel show the evolution of the period ratio of the middle to innermost  planet and
the upper right panel shows the evolution of the period ratio of the outermost  to  middle planet.
The curves marked with crosses are  obtained from  the analytic theory.  Equation (\ref{Anfit})  was used  as described in the text.
The central left panel shows the evolution of the angles $\Phi_1$ and $\Phi_2$  which  ultimately librate about $0$ and $\pi$ respectively.
 The central right panel shows the evolution of the angles $\Phi_3$ and $\Phi_4$  which  ultimately librate about $0$ and $\pi$ respectively.
The lower left and right panels show the evolution of the eccentricities on the middle and outermost planet respectively.
 \label{q10}}
\end{center}
\end{figure}


As values for the masses are not available, in order to estimate plausible values,  unless specified otherwise, 
we use the fiducial mass-radius relation  (see Lissauer et al. 2011) 
 \begin{equation} m/M_{\oplus} \propto (R_p/R_{\oplus})^{2.06}.\label{MRrel} \end{equation}
We use equation ({\ref{teccs}) to calculate the circularization time.
In order to implement this we need to specify the tidal parameter $Q'.$
We note that according to (\ref{teccs}), $t_{e,i}^s \propto Q' {\overline \rho}^{5/3},$
where ${\overline \rho_i}^{5/3}$ is the mean density of planet $i.$
Quoted standard deviations in the radii amount to  about 3 $\%$ for the inner two planets
and 6 $\% $ for the outermost planet. To within these limits, using the masses calculated from (\ref{MRrel}),  we may take 
the mean density,  ${\overline \rho_i} = 2.4\hspace{1mm}{\rm {g \hspace{1mm}cm^{-3}}}\equiv {\overline \rho},$ for each of the planets. 
Then,  if we  choose   $Q' = C_{Q'}({\overline \rho}/2.4)^{-5/3}$ for planet $i,$ with $C_{Q'}$ being a constant
we can scale our results to apply to hypothetical planets with the same masses  but with lower mean densities
and thus larger $Q'.$  Recall that $C_{Q'}\equiv Q'$ for the masses and radii  given in table \ref{table1}.

 The  simulations were  by means of $N$--body calculations  
using the method described in eg. Papaloizou \& Terquem (2001).    
 From considerations  of numerical tractability  we are unable  to consider very large values of $Q'.$ 
 In practice we took  either $C_{Q'}=2.5$ or $C_{Q'} =7.5$ for each planet.   However,  in one case considered below,  tidal dissipation was only applied to the 
 innermost planet.  Our general finding is that tidal dissipation  causes the system to evolve into a state where a strict  generalized Laplace resonance condition  holds
 with all resonant angles maintained in a state of libration. The period ratios then increase with time
with the system steadily separating  from strict commensurability. This type evolution is very similar to that obtained 
for two planet systems under similar circumstances (eg. Papaloizou 2011).
\subsection{Numerical Results}\label{Numressec}
\begin{table}
\begin{tabular}{|c|c|c|c|c| c|}
Planet &Radius&  Mass&  Mean density& Period&Period ratio \\
--&$R_p/R_{\oplus}$& $m/M_{\oplus}$&  ${ \overline \rho}$    &$P$&$(P_i/P_{i-1})$\\
1(b)&2.28&5.46&2.54 ${\rm{g \hspace{1mm} cm^{-3}}}$  &7.1316185&--\\
2(c)&2.47&6.44&2.36 ${\rm{g \hspace{1mm} cm^{-3}}}$ &8.9193459&1.2507\\
3(d)&2.55& 6.88&2.29\hspace{1mm}${\rm{g \hspace{1mm} cm^{-3}}}$  &11.9016171&1.3344\\
\end{tabular}
\caption{ The radii, estimated masses, estimated mean densities  and orbital periods in days for the three planets in the Kepler  60
system are given in the second to fifth columns respectively.
The sixth column  contains the period ratio with the next innermost planet where applicable. }
\label{table1}
\end{table}

\begin{figure}
\begin{center}
\epsfig{file=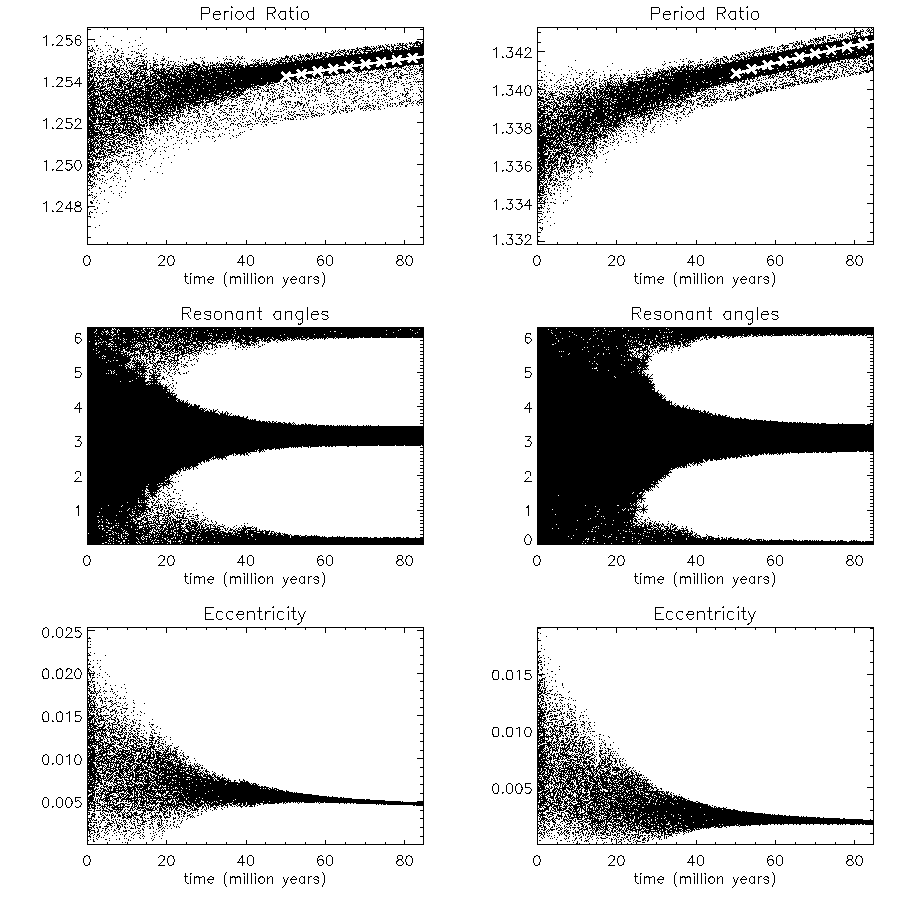
, width=16cm,height=18cm}
\caption{ As for  Fig. \ref{q10} but for  
 $Q'=7.5({\overline \rho}/2.4)^{-5/3}. $
  \label{q30}}
\end{center}
\end{figure}

The results obtained  for a simulation with $Q'~=~2.5({\overline \rho}/2.4)^{-5/3}$
are shown in 
 in Fig. \ref{q10}}. 
The  angles  $\Phi_1$ and $\Phi_2$ initially circulate  with the action of dissipation through tides  ultimately causing them to  librate with diminishing amplitude about $0$ and $\pi$ respectively.
 This evolution is well underway after a few million years.  If the simulation was performed without tidal dissipation,
 variations in all quantities would  remain at the same level with no tendency towards libration of the resonant angles.
 The angles $\Phi_3$ and $\Phi_4$ behave similarly to $\Phi_1$ and $\Phi_2,$   ultimately librating  about $0$ and $\pi$ respectively.
 This is all in accordance with the analytic model of section \ref{Lapmodel}.
The lower left and right panels of Fig. \ref{q10} show the evolution of the eccentricities on the middle and outermost planet respectively.
As the librational state is attained, the eccentricities approach slowly decreasing values with superposed small oscillations.
The period ratios of the middle to innermost  planet and
 the middle to  outermost planet then  secularly increase as expected from the analytic model  of section \ref{Lapmodel} for which the generalized three body Laplace relation is  assumed  to hold exactly.
 We  have plotted  expressions for the period ratios as a function of time 
obtained from  equation (\ref{Anfit})
 with  $1/ {\cal T}= 1.81\times 10^{-13}$\hspace{1mm}~yr$^{-1},$ 
being  the  value calculated from the analytic model, and
the  integration constant $t_0=3.5\times 10^6$~yr.
 We recall that in this case $q=4,$ and  the period ratios can be found from $|\Delta_{21}|$ by using
 $  16/5 (P_2/P_1 - 5/4) \sim |\Delta_{21}| /n_1$   and  then  the generalised three body  Laplace relation
 to obtain  the quantity $P_3/P_2 -4/3.$  
The curves marked with crosses  represent  these   expressions which track the numerical data quite well. 
 It is not clear exactly how the temporal fluctuations in the period
ratios should be averaged when considering  such fits. However, we  remark that the lightly shaded regions
 in the period ratio  plots are  found to be associated with high frequency perturbations 
 that are not represented in the analytic model.   On the other hand, 
the   darkly shaded  regions are associated with longer period librations.
Thus the fits  should apply to the  darkly shaded regions. We remark  that the above
analytically  determined curves  for $P_2/P_1$ and $P_3/P_2,$ as well as others discussed below,
obtained from equation  (\ref{Anfit}), were  calculated for positive values  of $t_0$ that are
comparable to the time for the resonant angles to begin circulating and which may be regarded
as a characteristic time for transient behaviour.
 However, adopting $t_0=0$ produces curves that appear
only slightly shifted from those presented.


 We also performed  a simulation with the same parameters except that 
tidal circularization applied only to
the innermost planet. The evolution is similar to that described for the previous case
but with the librational state taking longer to attain. We have  evaluated expressions for the period ratios as a function of time
obtained from  equation (\ref{Anfit}),
 for  $C_{Q'}=2.5.$ but with tidal effects acting only on the inner planet.
 For this case   we
found that the numerical results were reasonably well represented when 
 $1/ {\cal T}= 8.99\times 10^{-14}$~yr~$^{-1}$ as calculated from the analytic model,  together with  $t_0=3.5\times 10^6$~yr.
From this we see that the secular evolution of the librating state occurs at a rate that is a factor of two slower
when tides are applied only to the innermost planet. 
 
\subsubsection{Dependence on $Q'$}
In order to study the dependence of the evolution on $Q',$ we
show results  for  
$Q'=7.5({\overline \rho}/2.4)^{-5/3} ,$ with tidal effects applied to all the planets
in  Fig. \ref {q30}. In this case the evolution towards the librational state 
is as for the case with $Q'=2.5({\overline \rho}/2.4)^{-5/3} $
but three times slower as expected. This is apparent from the analytic   curves   obtained from equation (\ref{Anfit}).   
 Thus we found $1/ {\cal T}= 6.04 \times 10^{-14}~$yr$^{-1}$  and  adopted $t_0=1.05\times 10^7$~yr for this case.
In a  similar manner,  we have also confirmed the applicability of equation (\ref{Anfit})  for a run with $Q'=0.75({\overline \rho}/2.4)^{-5/3} ,$
 so that  the range of $Q'$ considered spans  one order of magnitude.  

\subsubsection{Dependence on planet mass}
We have explored the dependence of the evolution on planet mass by rerunning the  case  for which $Q'=7.5({\overline \rho}/2.4)^{-5/3} $
with all of the planet masses increased by a factor of two
while retaining a constant mean density. 
 The results were found to be qualitatively similar but the evolutionary time scale is shorter.
 We see from equation (\ref{teccs}) that the circularization time for each planet
scales as $m^{-2/3}_i$ and so will be reduced by a factor of $2^{2/3}.$
Equation (\ref{Delevfin}) then implies that ${\cal T}$ scales as $m^{-8/3}_i$ and so should  reduced by a factor $2^{8/3}.$
This is accurately confirmed by our   curves determined from  (\ref{Anfit})  that track the numerical
data. These   had   $1/ {\cal T}= 3.83\times 10^{-13}$~yr$^{-1}$  
  and $t_0=5.5\times 10^6$~yr.
We remark  that  if we had increased the planet masses by a  factor  of two while keeping their 
radii  and $Q'$ constant,  equations (\ref{teccs}) and (\ref{Delevfin}) would  imply that  ${\cal T}$ scales as $m^{-1}_i$ 
leading to a relatively  weaker reduction in the evolution time.
We also  attempted  simulations with the planet masses increased by a factor of four. However, these  were  found to lead to
disruption of the system through instability within a time $ < 10^6$~yr.
\begin{figure}
\begin{center}
\epsfig{file=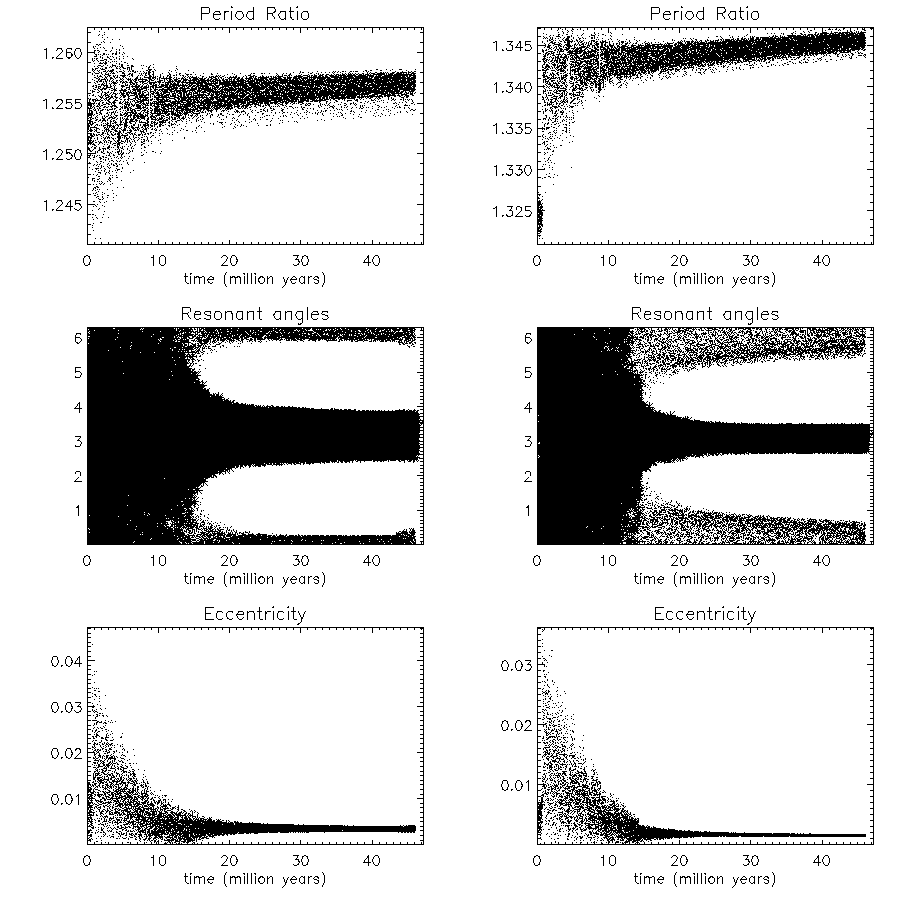
, width=16cm,height=18cm} 
\caption{As in Fig. \ref{q10} but with
different initial conditions described in the text.
 \label{q10shift}}
\end{center}
\end{figure}

\subsubsection{A small variation of the initial period ratios}
In this paper we have focused on initial conditions near to
those appropriate to the Kepler  60 system.
We remark that Marti et al. (2013) find that
the Laplace resonance in the non dissipative  GJ876 system
is  defined by a tiny island of regular motion surrounded by unstable highly chaotic orbits
( see also Gerlach \& Haghighipour 2012).
 We comment  that  we found  that some  runs
 starting with  slightly different initial conditions   led to the same  type of evolution as described above
 but  took significantly longer  to arrive at the  librating state.
We give results  for the case  with $C_{Q'}=2.5,$ but
with different initial conditions  to those used previously,  in
 Fig. \ref{q10shift}. These were such that the semi-major axis of the outermost planet
 was decreased by a factor of $1.0050,$ while the semi-major axis of the central planet 
 was decreased by a factor of $1.0007.$
 With these changes the initial value of the  period ratio $P_2/P_1$  changes from $1.2507$ to $1.2493$
 and  the initial value of the period ratio $P_3/P_2$ changes from $1.3344$ to $1.3259.$
 Accordingly these are both below the values required for strict commensurability.
 This simulation ultimately attains the same libration state, with secular increasing period ratios,
 as the case with the original initial conditions. However, the initial period of  time during  which 
the resonant angles circulate  and  there are relatively
 large eccentricity oscillations is seen to last about twice as long  at around $\sim 1.5\times 10^7$~yr  in this case.


\section{Discussion and conclusion}\label{sec6}
In this paper we have  constructed  a simple analytic model  that  describes the evolution of a 
three planet system in a three body resonance 
under the influence of tidal circularization.
This is based on adopting a time averaged system for which contributions from
a maximum of four resonant angles,   associated with the  first order resonances
between the inner pair and outer pair of planets, are retained.

When  all four angles undergo small amplitude libration  the generalized three body  Laplace relation  is maintained  
while the period ratios increasingly depart from strict commensurability.
But  when only three of the resonance angles  undergo small amplitude
libration, the generalized  three body Laplace relation was found to be  replaced  by a different relationship between
the period ratios which depends on details such as the planet mass ratios  and their  orbital circularization
time ratios.  

Features  of the analytic model  such as the scaling of the evolution time scale
with the  planet mass scale, its dependence on  tidal dissipation, and the form of the evolution of the
period ratios away from commensurabilty were reproduced 
by our numerical simulations which were carried out
for  a three planet system with masses in the super-earth regime that were focussed    
on the Kepler  60 system for which the inner pair exhibit a 5:4 resonance and the outer pair
exhibit a 4:3 resonance.  However, we remark that much of the discussion should be generic. 

We now briefly  discuss how  our results  can be use to provide constraints
on  the  amount of  tidal dissipation that has occurred in  such interacting systems.
This discussion  again  adopts working  parameters appropriate to the Kepler 60  system, though its  form should 
again be generic. 
 We  remark that although Steffen et al. (2013) confirmed the  Kepler 60 system by observing anti-correlated transit timing variations,
 low signal to noise parabolic,  rather than periodic,  variations are  seen.  Accordingly masses are not determined
and whether resonant angles are  in a librating  state is unclear.

 However, at the present time the inner and outer pair of planets are very close to 5:4 and 4:3 resonances respectively, with fractional  deviations
$\sim 10^{-3}.$ Such resonant chains are readily produced through convergent migration of protoplanets  driven by interaction with the gas disk.
Accordingly it is commonly assumed they were produced in this way while the gas disk was present (eg.
 Cresswell \& Nelson 2006; 
Terquem \& Papaloizou 2007;  Thommes et al. 2008;   Libert \& Tsiganis 2011;  Beaug\'e \& Nesvorny 2012, Cossou et al. 2013)
rather than through in situ formation, which tends to avoid resonances (Hansen \& Murray; 2013).
However, as noted by the former  authors such resonant chains may be broken  after the gas disperses  by 
dynamical  interactions and/or collisions  with remaining planetesimals.
Thus at a later time, but still shortly after formation,
the system may differ to some extent from a putative initial resonant liberating state formed through convergent migration.

Although the above scenario forms a plausible basis for the origin of planetary systems that are very close to comprising  a resonant 
chain, in principle the system could have started from a state that underwent  tidal evolution of the type described here, with low $Q',$
in such a way that the exact resonant state has been passed through only recently. However, noting that the evolution
is most rapid in the resonant state this would seem to be unlikely.
Accordingly, for the purposes of discussion, we shall  assume that  the system was formed close to the resonant state and 
 consider how it would subsequently evolve as a result of tidal
interactions with the central star.

 We begin by estimating  the time required for  the system to attain  a generalized  three body Laplace resonance with all four resonance angles
liberating with small amplitude. For simplicity in the discussion below, we assume that  same $Q'$ applies to all of the planets, noting
extension to consider more general cases should be straightforward.

As expected, the evolution time scales in our simulations have been found to scale with the tidal
dissipation parameter $Q'.$
From the results presented in Figs. \ref{q10} and \ref{q30}, the time required to attain a state in which  the four resonant
angles attain a state of small amplitude libration can be estimated to be   $\sim 6.7\times10^6 C_{Q'}$~yr.
Here we assume, as indicated by these results that  this time is  in general proportional to  $C_{Q'}.$ 
Recall  that $C_{Q'}\equiv Q'$ for  ${\overline \rho_i} = 2.4\hspace{1mm}{\rm {g \hspace{1mm}cm^{-3}}}$ as assumed here. 

Although they do not provide an estimate of the age of the system, the stellar parameters given 
by Batalha et al. (2013)  indicate that the star of mass $1.09 M_{\odot},$  radius  $1.5R_{\odot},$  and luminosity $2.5L_{\odot}$
is evolved.  Accordingly,  we shall adopt an age of $5\times 10^9$~yr  in order to make  illustrative estimates. 
Then  failure to attain  the  librating  state   implies that $Q'  > \sim 750.$
This applies if $Q'$  does not vary with time and   the same for all planets (but see below).  Should tides only  operate on the innermost planet,
the estimated bound on $Q'$ should be reduced by a factor of two.

 We now consider the expected evolution assuming that  the system   was formed in  the
four angle librating state.
Then we  note that if the system were actually in a three body  resonance with small amplitude librations,
the  observed  small deviations  from exact commensurability would indicate  a larger value of $Q'$ than is given by the above bound.
To show this, we consider  the evolutionary time scale,  $t_{ev},$  defined  as the characteristic time required for
$\Delta_{21}$ to  undergo a relative change of order unity.   Using (\ref{Delevfin1}), we find
\begin{equation}
t_{ev}= \left| \frac {\Delta_{21}}{d\Delta_{21}/dt}\right|=  3\left(\frac{\Delta_{21}}{n_1}\right)^3{\cal T}= 
\frac{12288}{125}\left( \frac{P_2}{P_1}-\frac{5}{4}\right)^3{\cal T}
\end{equation}
 Using  equations (\ref{Delevfin}) ond (\ref{Delevfin1}) while  noting
 the scaling that ${\cal T} \propto C_{Q'},$  we find  that ${\cal T}= 2.21\times 10^{12}C_{Q'}$~yr.  
Thus 
\begin{equation}
t_{ev}= 2.17\times10^{14}\left( \frac{P_2}{P_1}-\frac{5}{4}\right)^{3}C_{Q'}\hspace{1mm} {\rm yr}.\label{evolT0}
\end{equation}
 Adopting  the masses and  period ratios in table~\ref{table1} we then  obtain
\begin{equation}
t_{ev}= 7.44\times10^{4}C_{Q'} \hspace{1mm} {\rm yr}.\label{evolT}
\end{equation}
We recall that For the mean density ${\overline \rho} = 2.4\hspace{1mm}{\rm {g \hspace{1mm}cm^{-3}}}$  adopted, 
 $C_{Q'}\equiv Q'.$  In addition  for fixed  $Q',$ planetary  radii and planet  mass ratios, 
 $t_{ev}$ is inversely proportional
to the  planetary mass scale.
Equation (\ref{evolT}) 
  implies  that  the system   could not have begun  in a three body  resonance with four librating resonance angles and have 
its present period ratios
if $C_{Q'} <  \sim 6.7\times10^{4},$  this bound scaling with the mass scale. On the other hand if
 $C_{Q'} > \sim 10^{3}$   the  librating state  would not  have  been attained through circularization.

 Up to now we have assumed that $Q'$ is constant.
If $Q'$ varies with time, as evolution rates are $\propto (Q')^{-1},$ provided changes are slow enough
for the system to respond adiabatically, the above estimates should relate to $\langle (Q')^{-1} \rangle^{-1},$
where the angle brackets indicate a time average. This is effectively a statement that the amount of tidal dissipation
has to have  been limited.  Thus although  
periods of  strong episodic tidal heating of the type considered by Ojakangas \& Stevenson (1986)
may have occured, their integrated effect is constrained. 
We emphasise that the above estimates are provisional  and that it should be possible to  extend and refine  the above discussion as more information about  systems of this kind  becomes available. 
  

\end{document}